\documentclass[letter]{aa}

\usepackage{amsmath, amssymb, amsfonts, amscd}
\usepackage{mathtools} 
\usepackage{array} 
\usepackage[varg]{txfonts} 

\usepackage{graphicx} 

\usepackage{natbib}
\bibpunct{(}{)}{;}{a}{}{,} 

\usepackage[pdftex=true,pdffitwindow=false,plainpages=false,bookmarksopen=false,breaklinks=true,
pdfpagelabels=true,bookmarksnumbered=true,colorlinks=true, linkcolor=blue]{hyperref} 

\begin{document}

\title{Impact of the frequency dependence of tidal Q\\ on the evolution of planetary systems}

\author{
P. Auclair-Desrotour\inst{1,2,3}
\and C. Le Poncin-Lafitte\inst{1}
\and S. Mathis\inst{2,4}
}

\institute{SYRTE, Observatoire de Paris, UMR 8630 du CNRS, UPMC, 77 Av. Denfert-Rochereau, 75014 Paris, France
\and Laboratoire AIM Paris-Saclay, CEA/DSM - CNRS - Universit\'e Paris Diderot, IRFU/SAp Centre de Saclay, F-91191 Gif-sur-Yvette Cedex, France
\and IMCCE, Observatoire de Paris, UMR 8028 du CNRS, UPMC, 77 Av. Denfert-Rochereau, 75014 Paris, France
\and LESIA, Observatoire de Paris, CNRS UMR 8109, UPMC, Univ. Paris-Diderot, 5 place Jules Janssen, 92195 Meudon, France
\email{pierre.auclair-desrotour@obspm.fr;christophe.leponcin@obspm.fr;stephane.mathis@cea.fr} 
}

\date{Received ... / accepted ...}

\abstract
{Tidal dissipation in planets and in stars is one of the key physical mechanisms that drive the evolution of planetary systems.}
{Tidal dissipation properties are intrisically linked to the internal structure and the rheology of studied celestial bodies. The {resulting} dependence of the dissipation {upon} the tidal frequency is strongly different in the cases of solids and fluids.}
{{We compute the tidal evolution of a two-body coplanar system, using the tidal quality factor's frequency-dependencies appropriate to rocks and to convective fluids.}}
{{The ensuing orbital dynamics comes out smooth or strongly erratic, dependent on how
the tidal dissipation depends upon frequency.}}
{We demonstrate the strong impact of the internal structure and of the rheology of the central body on the orbital evolution of the tidal perturber. {A smooth frequency-dependence of the tidal dissipation renders a smooth orbital evolution while a peaked dissipation can furnish erratic orbital behaviour.}}


\titlerunning{Understanding tidal dissipation in stars and fluid planetary regions}
\authorrunning{Auclair-Desrotour, Mathis, Le Poncin-Lafitte}

\maketitle


\section{Introduction and context}

Tides are one of the key interactions that are driving the evolution of planetary systems. Indeed, because of the friction, both in the host-star and in planets' interiors, a system evolves either to a stable state of minimum energy, where spins are aligned, orbits are circularised, and the rotation of each body is synchronised with the orbital motion, or the perturber tends to spiral into the parent body \citep{Hut1980}. Therefore, understanding and modeling the dissipative mechanisms that convert the kinetic energy of tidally-excited velocities and displacements into heat is of great importance. These processes, driven by the complex response of a given body (either a star or a planet) to the gravific perturbation by a close companion, depends strongly on its internal structure and its rheology. Indeed, the tidal dissipation in solid (rocky/icy) planetary layers strongly differs from the dissipation in fluid regions in planets and in stars; the one in rocks and ices is often strong with a smooth dependence on the tidal frequency $\chi$, the one in gas and liquids being generally weaker in average and strongly resonant. Therefore, such properties must be taken into account in the study of the dynamical evolution of planetary systems using celestial mechanics.

To reach this objective, the tidal quality factor $Q$ has been introduced in the {literature} \citep{GS1966}. Its definition comes from {the evaluation of the tidal torque \citep{Kaula1964} and} the analogy with forced damped oscillators: it evaluates the ratio between the maximum energy stored in the tidal {distortion} during an orbital period and the energy dissipated by the friction. Indeed, a weak value of $Q$ corresponds to a strong dissipation and vice versa. In this framework, $Q$ can be computed from ab-initio resolution of the dissipative dynamical equations for the tidally-excited velocities and displacements in fluid and in solid layers of celestial bodies, respectively \citep[e.g.][]{Henningetal2009,Efroimsky2012,RMZL2012,Zahn1977,OL2004,OL2007,RMZ2012}. It leads to values of $Q$ that varies smoothly as a function of $\chi$ in rocks and ices while numerous and strong resonances are obtained in fluids. However, in celestial mechanics' studies, $Q$ is often assumed to be constant or to scale as $\chi^{-1}$ as convenient first approach and evaluated using scenario for the formation and the evolution of planetary systems.

In this work, we show how the dependence of $Q$ on $\chi$ impacts these evolution and must be taken into account. In Sect. 2, we describe the studied set-up and the corresponding dynamical equations, which correspond to the one adopted by \cite{EL2007} who studied the impact of the rheology of solids on related tidal dissipation and evolution (Sect. 3). In Sect. 4, we study the case of highly resonant tidal dissipation in fluid layers and discuss the strong differences with the case of solids. Finally, we discuss astrophysical consequences for the evolution of planetary systems.

\section{The studied set-up and the dynamical equations}

\subsection{The studied model} 

To study the impact of rheology on tidal evolution, and of the related variation of $Q$ as a function of $\chi$, we choose to follow the work by \cite{EL2007}. We thus study a two-body coplanar system with a central extended body A with a mass $M_{\rm A}$ and a mean radius $R_{\rm A}$ and a point-mass tidal perturber B of mass $M_{\rm B}$. In a reference fixed frame $ \mathcal{R}_{\rm A}: \left\{ A, \textbf{X}_{\rm A} , \textbf{Y}_{\rm A} , \textbf{Z}_{\rm A} \right\} $ the central-body is rotating with a spin vector $\vec\Omega_{\rm A}$ and the perturber is orbiting around it. In order to study a simplified system where we can easily isolate the effect of rheology, this motion is supposed circular. Thus, the position of B is directly given by the semi-major axis $ a $, which is the distance separating B from A in this particular case, and the mean anomaly $ \tilde{M}_{\rm B} =  n_{\rm B}t $, $ n_{\rm B} $ being the mean motion and $ t $ the time coordinate.

\subsection{Dynamical equations}

As recalled in the introduction, the tidal quality factor $Q$ is defined as the ratio between the maximum energy stored in the tidal distorsion during an orbital period and the energy dissipated by the friction. It is thus directly related to the rheology of studied bodies, that leads generally to a dependence of $Q$ as a function of the tidal frequency
\begin{equation}
\chi = \chi_{2200} = 2\,\vert n_{\rm B}-\Omega_{\rm A}\vert=\vert\omega_{2200}\vert=\vert\omega\vert
\end{equation}
\citep[e.g.][]{Greenberg2009,Efroimsky2012}, {where $\omega=\omega_{2200}$ is the principal, semidiurnal, Fourier tidal mode} that corresponds to the frequency of the perturbation in the frame rotating with the perturber. This friction induces a geometrical angle $\delta(\chi)$ between the directions of the tidal bulge and of the line of centers \footnote{{The following identity can be applied only to the $m=2$ case \citep[e.g.][]{EfroimskyMakarov2013}.}}
\begin{equation}
\delta\left(\chi\right)=\frac{1}{2} \chi {\Delta t}\left(\chi\right)=\frac{1}{2}\sin^{-1}\left[Q^{-1}\left(\chi\right)\right],
\end{equation}
where we have introduced the so-called time lag $\Delta t(\chi) $ \citep[see e.g.][]{Hut1981}.

This lag induces a net torque that modifies the evolution of the spin of body A \citep{MLP09}
\begin{equation}
\dfrac{d \Omega_{\rm A} }{dt} =  \frac{3}{2} \frac{k_2(\chi) G M_{\rm B}^2 R_{\rm A}^5}{I_{\rm A} a ^6} Q^{-1} \left( \chi \right) {\rm sgn} \left(\omega\right)\, ,
\label{dynspin}
\end{equation}
where $ I_{\rm A} $ is the moment of inertia of body A, $k_2(\chi)$ is the Love number and $ G $ the gravitational constant. The semi-major axis of body B is also modified by \citep{EL2007}
\begin{equation}
\dfrac{da}{dt} = - \frac{3 k_2(\chi) R_{\rm A}^5 n_{\rm B} M_{\rm B}}{M_{\rm A} a^4 } Q^{-1}\left(\chi\right) {\rm sgn} \left(\omega\right).
\label{dynorb}
\end{equation} 
These equations show that $ Q^{-1} $ has explicitly a linear impact on the evolution of the system; strong variations of $ Q^{-1} $ thus imply rapid changes for $ a $ and $\Omega_{\rm A}$. Then in this study we take into account the dependence of $Q$ to $\chi$ \citep[e.g.][]{MLP09,EfroimskyMakarov2013} in order to consider straightforwardly the impact of rheological models.

\section{The case of solid tides}

To solve our problem, we must close it with the choice of a law giving $Q$ {(or more generally, $k_{2}/Q$)} as a function of $\chi$. {It is common to assume $Q$ to be either} constant \citep[e.g.][]{1964RvGSP...2..467M} or to scale as $\chi^{-1}$ in the case of a constant tidal time lag \citep[e.g.][]{Hut1981}. However, for solid rocky or icy bodies, \cite{EL2007} suggest to use a power scaling law\footnote{{Here, we neglect the frequency-dependence of the Love number. For realistic materials, the latter approximation is legitimate at frequency much higher than the inverse Maxwell time.}}: $ Q = \mathcal{E}^{\alpha}\chi^{\alpha} $, which has been experimentaly validated {for} metals and silicates {in the lab, as well as in seismic and geodetic experiments}. Here, the empirical parameters $ \alpha $ and $ \mathcal{E} $ are bound to the rheology; $ \alpha $ characterises the frequency dependence and takes values between $ 0.1 $ and $ 0.4 $; $ \mathcal{E} $ is {an integral relaxation parameter, which has the dimension of time and is determined by the internal-friction mechanism dominating at the frequency $\chi$. Usually, one or another mechanism or group of mechanisms stay dominant over vast bands of frequencies. Over these bands, $\mathcal{E}$ may be regarded constant or almost constant.} This law is particularly interesting since it introduces the frequency dependence with only one more parameter than the constant law and keeps close to the realistic physics of solids at the same time.

To compute the evolution of $ a $ and $ \Omega_{\rm A} $ with time, we {chose} to use the numerical code developed by one of us and based on the code ODEX \citep{hairer1993}. In order to validate it, we studied the case of the system Mars-Phobos simulated by \cite{EL2007} assuming exactly the same parameters that are summarised in table \ref{para_simu_phobos}. The Phobos' initial semi-major axis, the initial dissipative time lag and constant tidal quality factor are respectively denoted $ a_0 $, $ \Delta t_0 $ and $Q$.

\begin{table}[h]
\centering
    \begin{tabular}{ c c }
      \hline
      \hline
      \textsc{Parameters} & \textsc{Numerical values} \\
      \hline
      $ G $ & $ 6.67384.10^{-11} $ $ {\rm m^3 kg^{-1} s^{-2} }$ \\
      $ M_A $ & $ 6,4185.10^{23} $ $ {\rm kg} $ \\
      $ R_A $ & $ 3,3962.10^{3} $ $ {\rm km} $ \\
      $ \Omega_{\rm A} $ & $ 7,08822.10^{-5} $ $ {\rm rad.s^{-1}} $ \\
      $ k_2 $ & $ 0,152 $ \\
      $ Q $ & $ 79,91 $ \\
      $ M_B $ & $ 1,0189.10^{16} $ $ {\rm kg} $ \\
      $ a_0 $ & $ 9,3771.10^3 $ $ {\rm km} $ \\
      $ \Delta t_0 $ & $ 39,864 $ $ {\rm s} $ \\
      \hline
    \end{tabular}
    \textsf{\caption{\label{para_simu_phobos} Numerical values used in the simulation of the system Mars Phobos.}}
\end{table}
 
We note that in this case, the spin of A does not change over time compared to $ a $ given that:
\begin{equation}
\left| \frac{d \Omega_{\rm A}}{\Omega_{\rm A}}.\frac{a}{da} \right| = \frac{G M_{\rm A} M_{\rm B}}{2 \Omega_{\rm A} n_{\rm B} a I_{\rm A}} \approx 10^{-7} \ll 1.
\end{equation}

\begin{figure}[htb]
\centering
{\includegraphics[width=0.325\textwidth]
{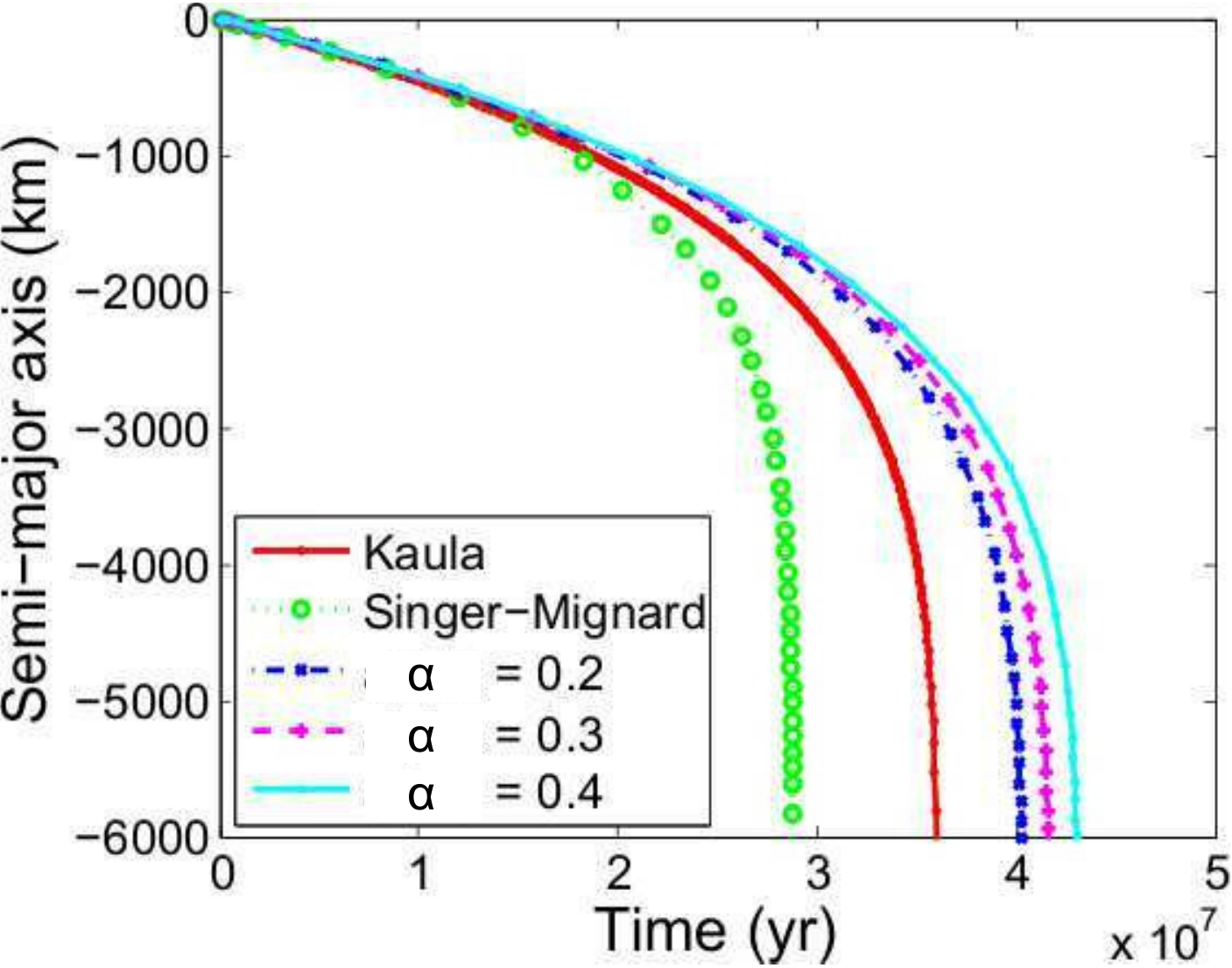}
\textsf{\caption{\label{fig:courbes_a} Temporal evolution of the semi-major axis $ a $ {for a class of tidal models with $k_2/Q\propto\chi^{-\alpha}$ and with} different values of the parameter $ \alpha $. {The abscissa representing time in years, the vertical axis measures the deviation of the semi major-axis from its initial value $a_0$.} Our plots serve to compare realistic rheologies ($\alpha=0.2,0.3,0.4$) with less physical ones - those with $\alpha=0$ \citep{Kaula1964} and $\alpha=-1$  \citep{Singer1968,Mignard1979}. }}}
\end{figure}

Our results perfectly reproduce those obtained by \cite{EL2007}. Figure \ref{fig:courbes_a} shows the evolution of $ a $ with time for different values of $ \alpha $ = $ -1 $ (constant tidal time lag $\Delta t_0$; e.g. \cite{Singer1968,Mignard1979}), $0$ (constant $Q$; e.g. \cite{Kaula1964}), $ 0.2 $, $ 0.3 $, and $ 0.4 $. These plots already highlight the impact of the rheology on the smooth induced evolution of orbital parameters such as the semi-major axis and on the related life-time of the system. 

\section{The case of tidal dissipation in fluid layers}

\subsection{Frequency dependence of Q} 

Revisiting the work by \cite{EL2007} provides us a strong basis to explore the impact of tidal dissipation in fluid bodies. In this section, we thus choose to study the evolution of a perturber of the mass of Phobos orbiting around a hypothetical completely fluid central body with the mass of Mars. Therefore, only the tidal quality factor $Q\left(\chi\right)$ will change.\\ 

In fluid bodies, tidal dissipation is due to the turbulent viscous friction acting on the equilibrium tide and on inertial waves, which are driven by the Coriolis acceleration, in convective regions \citep[e.g.][]{Zahn1977,OL2004,OL2007,RMZ2012} and on thermal and viscous diffusions acting on gravito-inertial waves in stably stratified zones \citep[e.g.][]{Zahn1977,OL2004,OL2007}. The excitation of this eigenmodes of oscillation by tides then leads to a highly resonant dissipation. 

From now on, to illustrate our purpose, we consider that the central body is completely convective and rapidly rotating {so} that $0\le\sigma\le1$, where $\sigma=\chi/\left(2\Omega_{\rm A}\right)$, giving birth to tidally-excited inertial waves. There is then a strong difference between the tidal quality factor adopted before for solid bodies that scale as a smooth power-law of $\sigma$ and the one related to inertial waves. Indeed, as demonstrated by \cite{OL2004}, using a local approach, their viscous dissipation is expressed as a sum of corresponding resonant terms\footnote{Global models lead to the same behaviour.}
\begin{equation}
D\!\left(\sigma\right)\!=\!D_0 \!\! \sum_{\left\{m,n\right\} \in \mathbb{N^{*}}\times\mathbb{N^{*}}} \!\! \frac{(m^2 + n^2) \left| \tilde{\sigma}^2 \right| + n^2}{ \left| (m^2+n^2)\,\tilde{\sigma}^2 - n^2 \right|^2 } \left(  m^2 + n^2 \right) \left| n f_{mn} - m h_{mn} \right|^2\!\!,
\label{DI}
\end{equation}
where $ \tilde{\sigma} = \sigma + i E \left( m^2 + n^2 \right) $ and $ E= \nu /\left(2 \Omega_A L^2\right) $ is the Ekman number of the fluid, $ \nu $ being the viscosity and $ L $ a characteristic length; $m$ and $n$ are the vertical and horizontal wave-vectors of inertial waves, respectively; finally, $ f_{mn} $ and $ h_{mn} $ are the coefficients of the Fourier series of the excitation. The tidal dissipation is thus a complex set of resonant peaks depending on the viscosity and on the rotation of the fluid. Since $ Q\left(\sigma\right)\propto\left[D\left(\sigma\right)\right]^{-1}$, we coupled it with the dynamical equations Eqs. (\ref{dynspin}-\ref{dynorb}) of our model.

\subsection{Numerical integration}

\begin{figure}[h!]
\centering
{\includegraphics[width=0.36\textwidth]{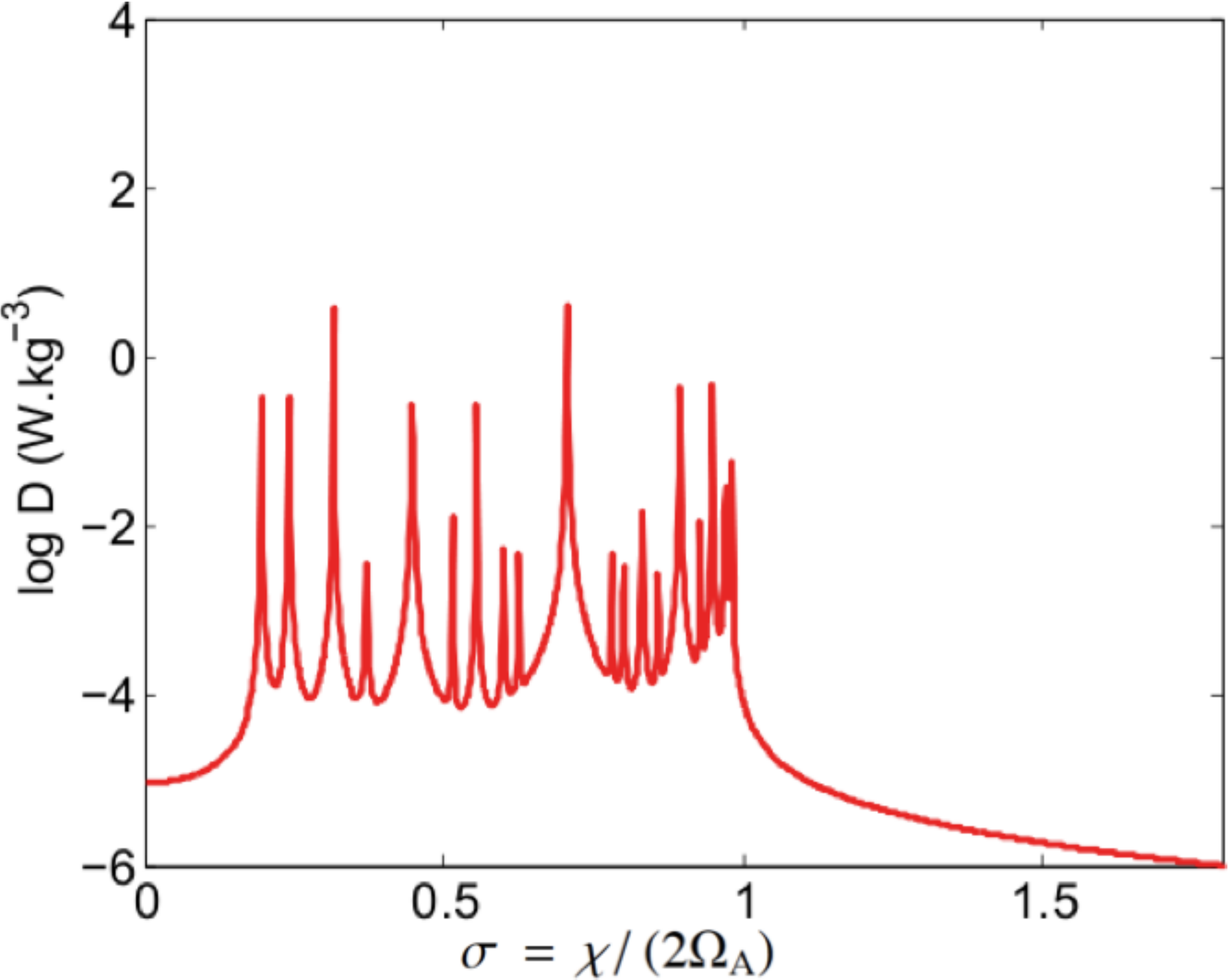}
\textsf{\caption{\label{fig:test_D} Resonant tidal dissipation spectrum {\bf $D$} resulting from inertial modes as a function of the normalised tidal frequency $\sigma=\chi/(2\Omega_{\rm A})$} assuming $ N_{\rm max} = 5 $. }}
\end{figure}

To evaluate the effects of such resonances on dynamics, we thus compute the evolution of the semi-major axis of the orbit with the same parameters that in the case of solid tides but giving as input a synthetic $Q^{-1}\left(\sigma\right)$ factor written like $D\left(\sigma\right)$ given in Eq. (\ref{DI}). Our fluid is characterised by its Ekman number, $ E = 10^{-5} $, which is a value often adopted in the literature for planetary convective layers and that allows to get a peaked dissipation (see Fig. \ref{fig:test_D}) \footnote{The Eckman number depends on the modeling of the turbulent viscosity  \citep[e.g.][]{OL2012}.}. The maximal rank of the sum ($N_{\rm max}$) is chosen to be relatively low, with $ N_{\rm max} = 5 $, in order to increase the speed of computation. Following \cite{OL2004}, we describe the excitation with the coefficients
\begin{equation}
\begin{array}{cccc}
f_{mn} = \displaystyle{\frac{1}{mn^2}}, & g_{mn} = 0, & \mbox{and} & h_{mn} = 0. 
\end{array}
\end{equation}

\begin{figure}[h!]
\centering
{\includegraphics[width=0.32\textwidth]{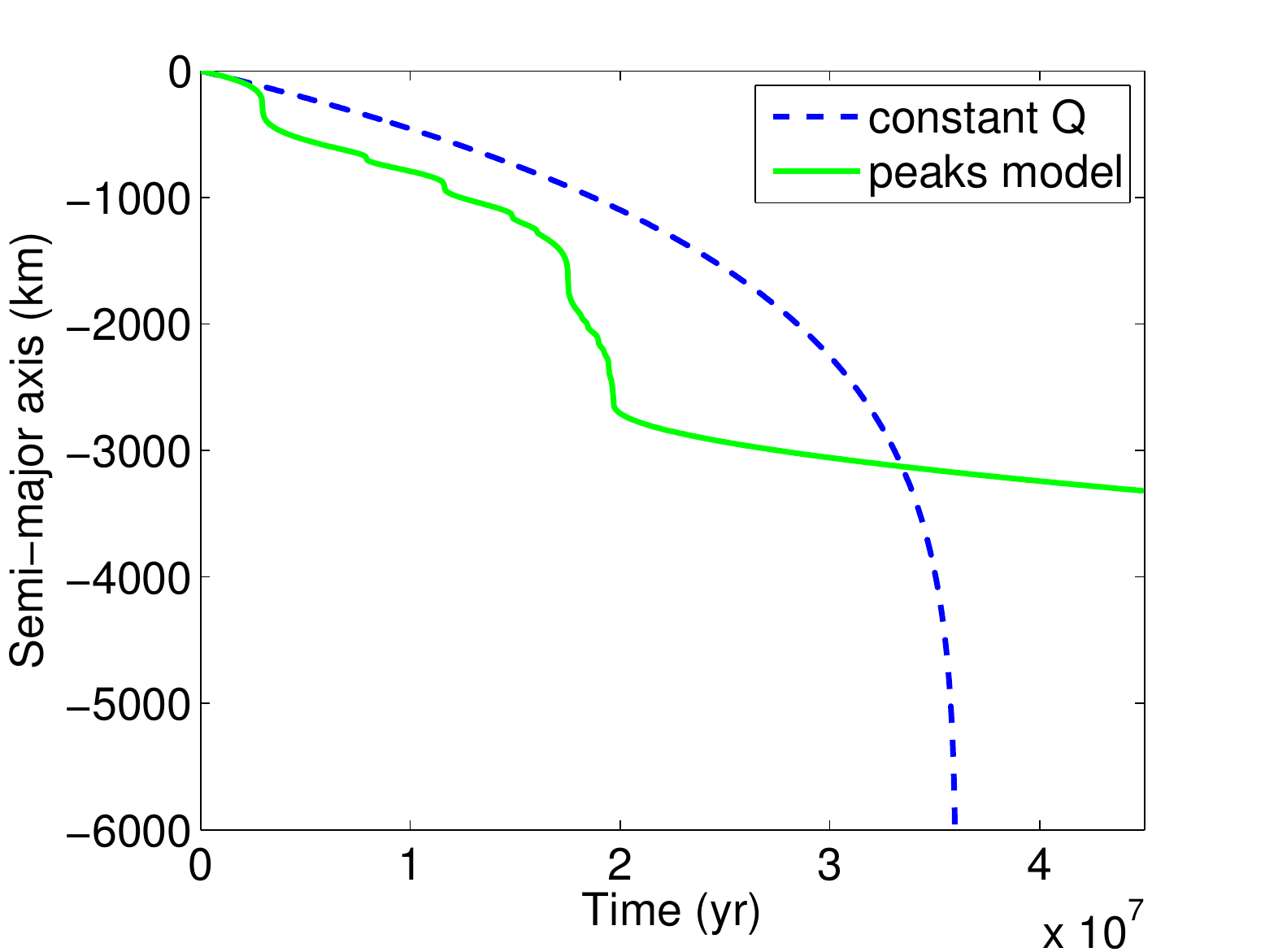}
\textsf{\caption{\label{fig:test_a} Evolution of the semi-major axis $ a $ over time with a $ Q $ factor proportional to inertial waves dissipation in fluids (green curve), and with a constant $ Q $ factor (blue dashed curve). The abscissa represents time in years, the vertical one measures the evolution of the semi major-axis from the initial value $a_0$.} }}
\end{figure}

The simulation clearly shows that, contrary to the case of solid tides, where the power scaling law implies a smooth evolution of $ a $, a contrasted $ Q $ factor, drastically depending on the tidal frequency, gives place to abrupt changes of $a$ (see Fig. \ref{fig:test_a}). As the perturber comes nearer from the central body, its mean motion increases. So dissipation strongly varies along the evolution of the system and, at each time it meets a resonance, there is a jump of $ a $, which is bounded to the properties of the peak: the higher and wider the peak, the greater the amplitude of the jump. This is the resonance-locking identified by \cite{WS1999} in the stellar case. However, we note that when $\sigma >1$, at the end of the simulation, the evolution of $a$ becomes smooth again. The reason for this behaviour is that we are outside the range of frequencies where inertial waves are excited. Then, the tidal dissipation is the one of the equilibrium tide that corresponds to the non-resonant background of $D$ observed in Fig. \ref{fig:test_D}. Finally, as demonstrated in Fig. \ref{fig:test_a}, the evolution of a system where the tidal dissipation is due to resonant eigenmodes (here the inertial waves) cannot be described properly using models where $Q$ or $\Delta t$ are assumed to be constant (respectively $\alpha=0$ and $\alpha=-1$).


\subsection{Scaling law}

\begin{figure*}[!t]
\begin{center}
   \includegraphics[width=0.24\linewidth]{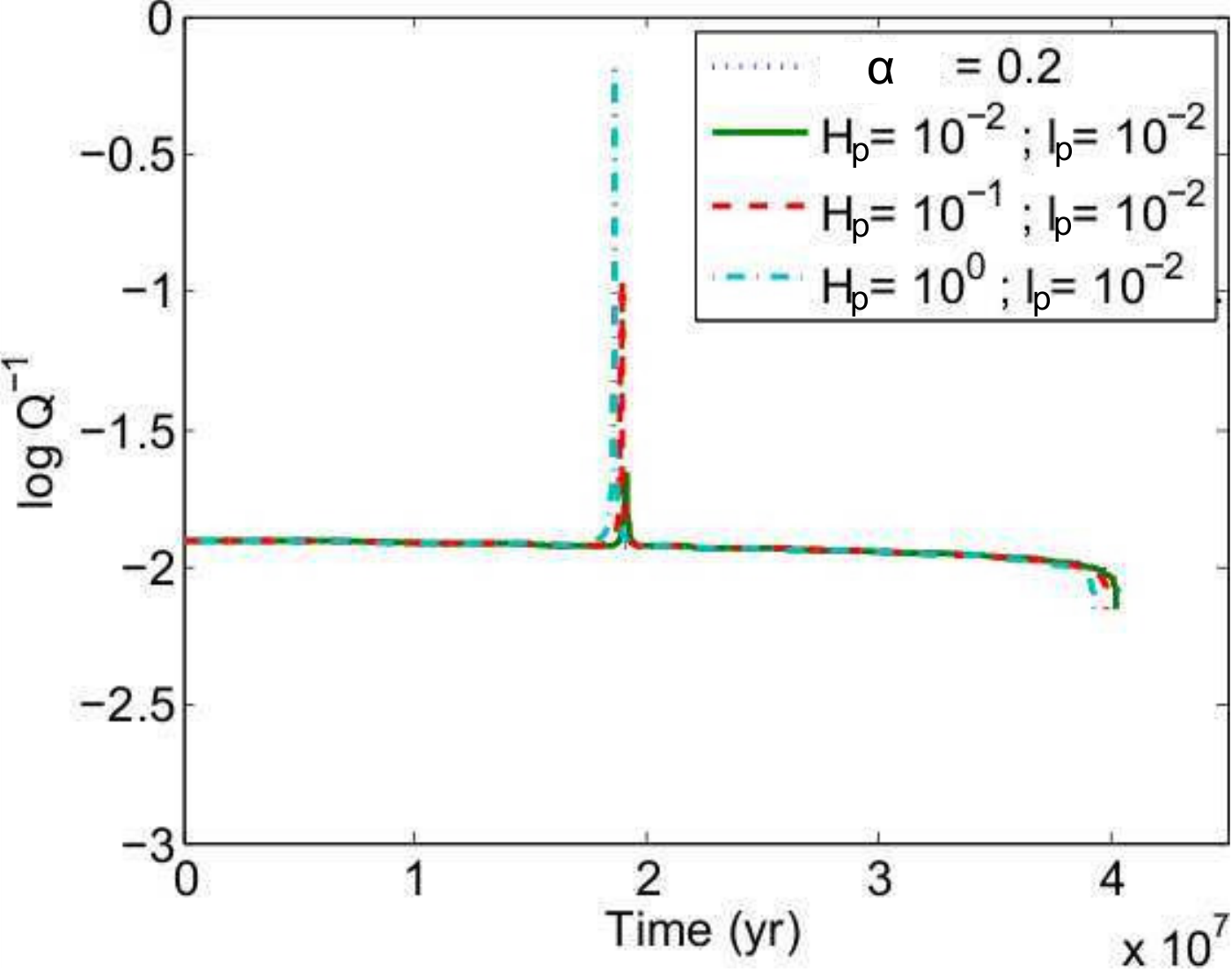}
   \includegraphics[width=0.24\linewidth]{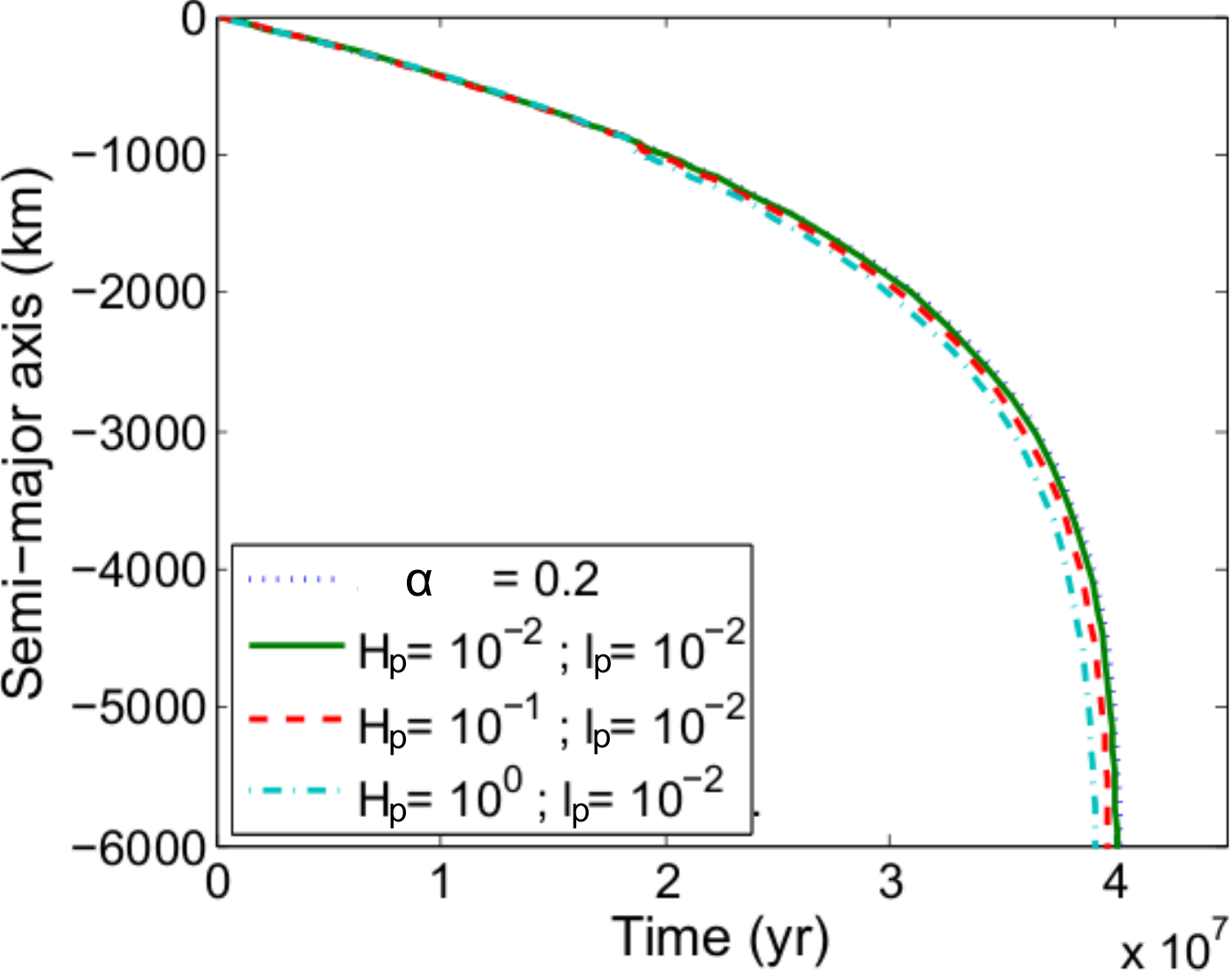}     
    \includegraphics[width=0.24\linewidth]{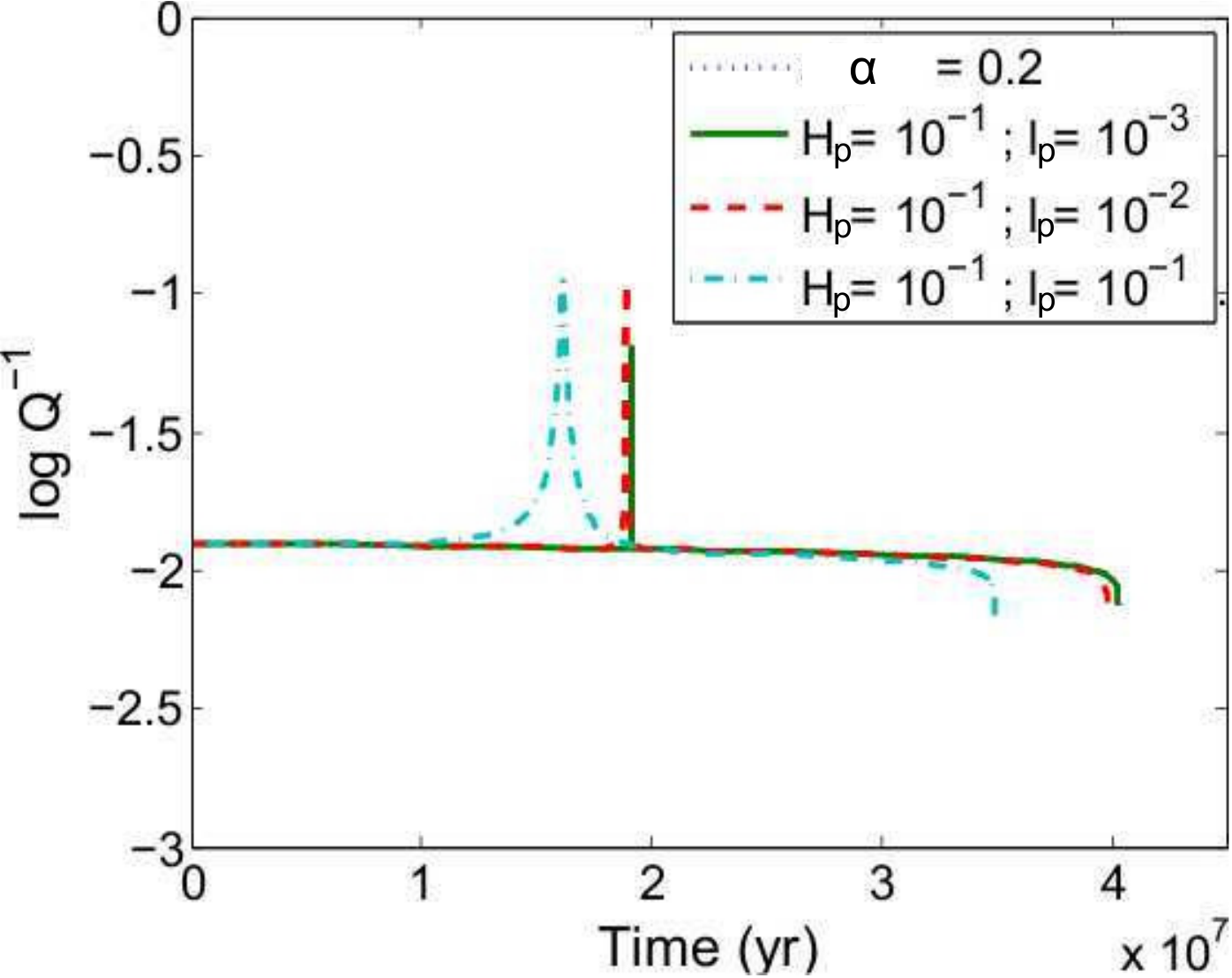}  
   \includegraphics[width=0.24\linewidth]{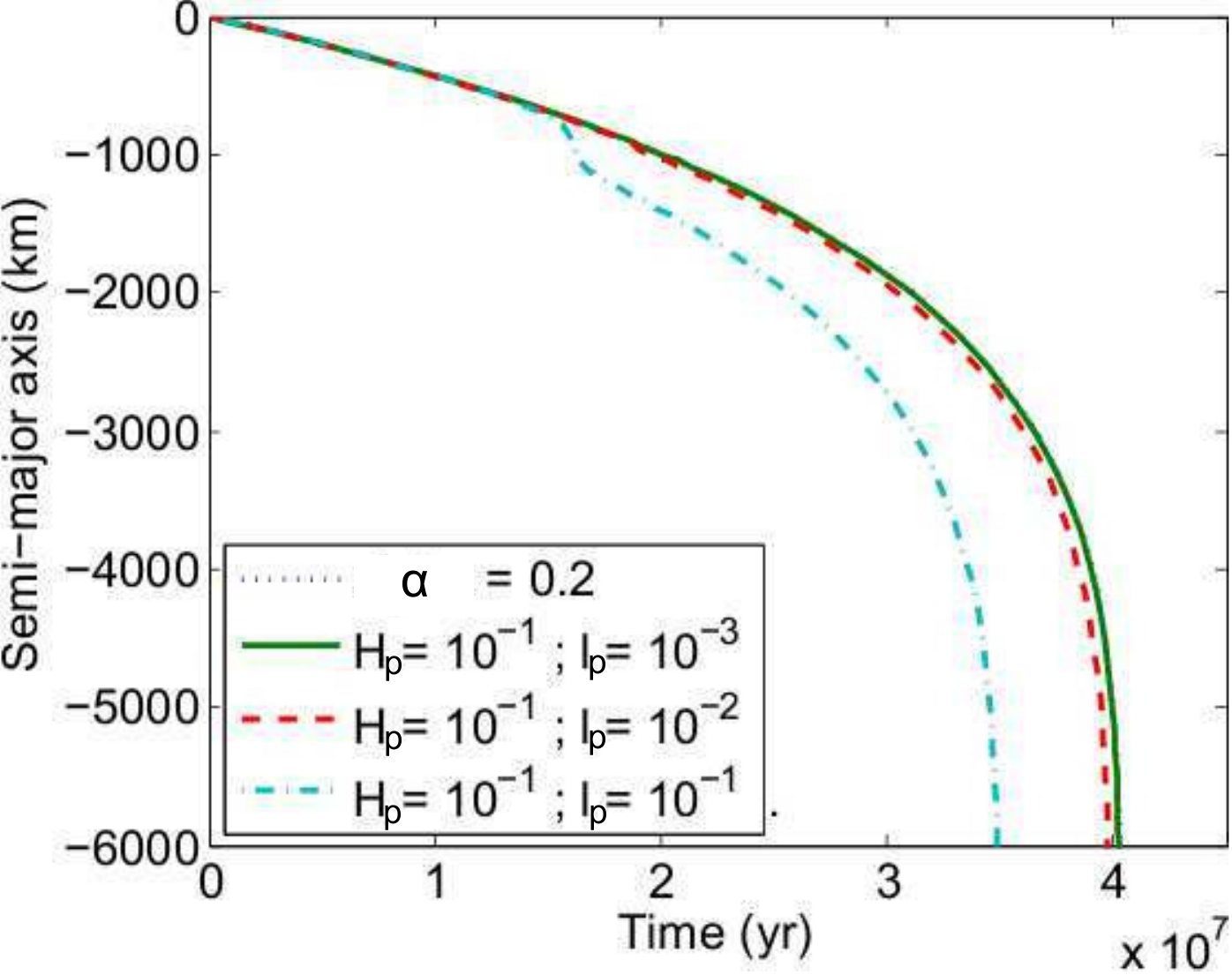}     
\end{center}
  \caption{\label{fig:pics_H_l} Evolution of the tidal dissipation and of the semi-major axis with time for different values of $ l_p $ and $ H_p $, the width at half-height and the height of the studied single resonant damping peak. The grey dotted line corresponding to $\alpha=0.2$ is superposed to the continuous green one except at the position of the peak for $Q^{-1}$.}
\end{figure*}

The resonant properties of the tidal dissipation in fluids (see Fig \ref{fig:test_D}) give birth to jumps of the value of the semi-major axis $a$ during the evolution of the system (see Fig. \ref{fig:test_a}). In this framework, the link between the orbital dynamics and the rheology of the fluid is the shape of resonances of the viscous dissipation $D$ (see Eq. \ref{DI}). A resonance occurs when a term of the sum in Eq. (\ref{DI}) becomes greater than all the others. In this section, our goal is thus to obtain a scaling law relating a jump of $a$ to the height, $H_p$, and to the width at half-height, $l_p$, of the corresponding single resonant damping peak (see Fig. \ref{fig:pics_H_l}) defined as
\begin{equation}
Q_{\rm p}^{-1} \left( \sigma \right) = \frac{H_{\rm p}}{ \left[ 4 \left( \sqrt{2} -1 \right) \left(\displaystyle{ \frac{ \sigma - \sigma_{\rm p} }{l_{\rm p}} } \right)^2 +1 \right]^2 },
\label{Qp}
\end{equation}
where $ \sigma_p $ is the resonant frequency. Then, the dissipation $ Q^{-1} $ is chosen to be the sum of a smooth background denoted $ Q_0^{-1} $ that corresponds to the one studied in \S 3. and of a resonant one $ Q_{\rm p}^{-1} $ (Eq. \ref{Qp}) that leads to the following equation for $a$ using Eq. (\ref{dynorb})
\begin{equation}
\dfrac{da}{dt} = - \frac{3 k_2 R_A^5 n_B M_B}{M_A a^4 } \left[ Q_0^{-1} \left( \sigma \right) + Q_p^{-1} \left( \sigma \right) \right]  {\rm sgn} (\omega).
\end{equation} 

Supposing that the peak has an influence on the system when the condition $ Q_p^{-1} \geq Q_0^{-1} $ is fullfilled, and that the resulting variation is rapid compared to the mean evolution, we can derive the amplitude of the jump
\begin{equation}
\frac{\Delta a}{a} \approx \frac{2 l_{\rm p}}{ 3 \sqrt{ \sqrt{2} -1 } \left( 1 + \sigma_{\rm p} \right)} \left[ \sqrt{ \frac{H_{\rm p}}{ Q_0^{-1} \left( \sigma_{\rm p} \right) } } - 1  \right]^{\frac{1}{2}}.
\label{SL}
\end{equation}
In Fig. \ref{fig:pics_H_l}, we plot the evolution of the semi-major axis for different values of $ H_{\rm p} $ and $ l_{\rm p} $ and the corresponding dissipation. These graphs illustrate the scaling law (Eq. \ref{SL}) by showing that the width of a peak has a greater impact on $ a $ than its height. Moreover, the values of $\Delta a/a$ obtained using direct numerical simulations perfectly match with those predicted by Eq. (\ref{SL}). Finally as $\sigma_{\rm p}$, $H_{\rm p}$, $l_{\rm p}$ are directly related to the value of the Eckman number $ E= \nu /\left(2 \Omega_{\rm A} L^2\right) $ \citep[c.f.][]{OL2004}, we see how the orbital dynamics is directly impacted by the fluid rheology and resonances.



\section{Conclusions}

In this work, we examined the impact of the frequency dependence of tidal dissipation in solids and fluids on the orbital evolution of a coplanar two body system. We show the strong different evolutions induced by tides in rocks and by tides exerted on fluid layers where eigenmodes are resonantly excited. A smooth dependence of the tidal dissipation on the tidal frequency drives a smooth orbital evolution while a peaked dissipation induces an erratic one. In each case, we point the direct impact of the rheology's properties on the dynamics of the system. Finally, this work shows how it becomes important to take the dependence of the tidal dissipation on the tidal frequency into account and the important consequences it may have for the evolution of star-planet(s) and planet-moon(s) systems in the Solar and exoplanetary systems. {In this context, the impact of the frequency dependence of the tidal torque on resonances will be examined in a forthcoming work.}\\





\begin{acknowledgements}
{The authors are grateful to the referee, M. Efroimsky, for his detailed review which has allowed to improve the paper.} P. Auclair Desrotour was supported by the Paris Observatory (SYRTE). This work was supported by the Programme National de Plan\'etologie (CNRS/INSU), the GRAM specific action (CNRS/INSU-INP, CNES), the Paris Observatory, the Campus Spatial de l'Universit\'e Paris Diderot, and the Emergence-UPMC grant (contract number: EME0911). {C.L.P.L. and S. M. dedicate this article to Dr. M. Le Poncin.}
\end{acknowledgements}

\bibliographystyle{aa}  
\bibliography{ADLPMref} 

\begin{thebibliography}{21}
\expandafter\ifx\csname natexlab\endcsname\relax\def\natexlab#1{#1}\fi

\bibitem[{{Efroimsky}(2012)}]{Efroimsky2012}
{Efroimsky}, M. 2012, \apj, 746, 150

\bibitem[{{Efroimsky} \& {Lainey}(2007)}]{EL2007}
{Efroimsky}, M. \& {Lainey}, V. 2007, Journal of Geophysical Research
  (Planets), 112, 12003

\bibitem[{{Efroimsky} \& {Makarov}(2013)}]{EfroimskyMakarov2013}
{Efroimsky}, M. \& {Makarov}, V.~V. 2013, \apj, 764, 26

\bibitem[{{Goldreich} \& {Soter}(1966)}]{GS1966}
{Goldreich}, P. \& {Soter}, S. 1966, \icarus, 5, 375

\bibitem[{{Greenberg}(2009)}]{Greenberg2009}
{Greenberg}, R. 2009, \apjl, 698, L42

\bibitem[{Hairer {et~al.}(2000)Hairer, N{\o}rsett, \& Wanner}]{hairer1993}
Hairer, E., N{\o}rsett, S., \& Wanner, G. 2000, Solving Ordinary Differential
  Equations {I} Nonstiff problems, 2nd edn. (Berlin: Springer)

\bibitem[{{Henning} {et~al.}(2009){Henning}, {O'Connell}, \&
  {Sasselov}}]{Henningetal2009}
{Henning}, W.~G., {O'Connell}, R.~J., \& {Sasselov}, D.~D. 2009, \apj, 707,
  1000

\bibitem[{{Hut}(1980)}]{Hut1980}
{Hut}, P. 1980, \aap, 92, 167

\bibitem[{{Hut}(1981)}]{Hut1981}
{Hut}, P. 1981, \aap, 99, 126

\bibitem[{{Kaula}(1964)}]{Kaula1964}
{Kaula}, W.~M. 1964, Reviews of Geophysics and Space Physics, 2, 661

\bibitem[{MacDonald(1964)}]{1964RvGSP...2..467M}
MacDonald, G.~J.~F. 1964, Reviews of Geophysics and Space Physics, 2, 467

\bibitem[{{Mathis} \& {Le Poncin-Lafitte}(2009)}]{MLP09}
{Mathis}, S. \& {Le Poncin-Lafitte}, C. 2009, \aap, 497, 889

\bibitem[{{Mignard}(1979)}]{Mignard1979}
{Mignard}, F. 1979, Moon and Planets, 20, 301

\bibitem[{{Ogilvie} \& {Lesur}(2012)}]{OL2012}
{Ogilvie}, G.~I. \& {Lesur}, G. 2012, \mnras, 422, 1975

\bibitem[{{Ogilvie} \& {Lin}(2004)}]{OL2004}
{Ogilvie}, G.~I. \& {Lin}, D.~N.~C. 2004, \apj, 610, 477

\bibitem[{{Ogilvie} \& {Lin}(2007)}]{OL2007}
{Ogilvie}, G.~I. \& {Lin}, D.~N.~C. 2007, \apj, 661, 1180

\bibitem[{{Remus} {et~al.}(2012{\natexlab{a}}){Remus}, {Mathis}, \&
  {Zahn}}]{RMZ2012}
{Remus}, F., {Mathis}, S., \& {Zahn}, J.-P. 2012{\natexlab{a}}, \aap, 544, A132

\bibitem[{{Remus} {et~al.}(2012{\natexlab{b}}){Remus}, {Mathis}, {Zahn}, \&
  {Lainey}}]{RMZL2012}
{Remus}, F., {Mathis}, S., {Zahn}, J.-P., \& {Lainey}, V. 2012{\natexlab{b}},
  \aap, 541, A165

\bibitem[{{Singer}(1968)}]{Singer1968}
{Singer}, S.~F. 1968, Geophysical Journal of the Royal Astronomical Society,
  15, 205

\bibitem[{{Witte} \& {Savonije}(1999)}]{WS1999}
{Witte}, M.~G. \& {Savonije}, G.~J. 1999, \aap, 350, 129

\bibitem[{{Zahn}(1977)}]{Zahn1977}
{Zahn}, J.-P. 1977, \aap, 57, 383

\end{thebibliography}

\end{document}